\documentclass[prb,twocolumn,showpacs,preprintnumbers]{revtex4}

\usepackage{graphicx}
\usepackage{amsfonts}
\usepackage{amssymb}
\usepackage{amsmath}
\usepackage{bm}

\newcommand{\be}{\begin{equation}}
\newcommand{\ee}{\end{equation}}
\newcommand{\bea}{\begin{eqnarray}}
\newcommand{\eea}{\end{eqnarray}}
\newcommand{\Eq}[1]{Eq.\,(\ref{#1})}
\newcommand{\Fig}[1]{Fig.\,\ref{#1}}
\newcommand{\Sec}[1]{Sec.\,\ref{#1}}
\newcommand{\Onlinecite}[1]{Ref.~\onlinecite{#1}} 

\newcommand{\epss}{\epsilon_s}
\newcommand{\epsC}{\epsilon_C}
\newcommand{\epsH}{\epsilon_H}
\newcommand{\epsL}{\epsilon_L}
\newcommand{\omC}{\omega_C}
\newcommand{\kappaC}{\varkappa_C}
\newcommand{\etaL}{\eta_L}
\newcommand{\etaH}{\eta_H}
\newcommand{\etaC}{\eta_C}
\newcommand{\thetaC}{\theta_C}
\newcommand{\lambdaC}{\lambda_C}
\newcommand{\LC}{L_C}
\newcommand{\nC}{n_C}

\makeatletter
\newcommand\res{\mathop{\operator@font {\rm Res}}\nolimits}
\makeatother

\begin{document}

\title{Resonant state expansion applied to planar waveguides}

\author{L. J. Armitage}
\author{ M.\,B. Doost}
\author{ W.\, Langbein}
\author{E.\,A. Muljarov}
\altaffiliation[]{egor.muljarov@astro.cf.ac.uk} 
\affiliation{School of Physics and Astronomy, Cardiff University, Cardiff CF24 3AA,
United Kingdom}
\begin{abstract}

The resonant state expansion, a recently developed method in electrodynamics, is generalized here to planar open optical systems with non-normal incidence of light. The method is illustrated and verified on exactly solvable examples, such as a dielectric slab and a Bragg reflector microcavity, for which explicit analytic formulas are developed. This comparison demonstrates the accuracy and convergence of the method. Interestingly, the spectral analysis of a dielectric slab in terms of resonant states reveals an influence of waveguide modes in the transmission. These modes, which on resonance do not couple to external light, surprisingly do couple to external light for off-resonant excitation.

\end{abstract}


\date{\today}

\maketitle

\section{Introduction}

Optical waveguides (WGs) are a basic building block for optical technology owing to their lossless guiding of light, enabled for example by total internal reflection. WGs provide confinement of the light in one or two dimensions, while allowing light waves to propagate along the remaining dimensions in which the waveguides are approximately invariant.  Planar WGs with one-dimensional (1D) confinement, such as a dielectric slab, and fiber WGs with  two-dimensional (2D) confinement, are widely used, for example in fiber optic cables for telecommunication, photonic crystal fibers,\cite{Knight2003} integrated optical circuits\cite{John2009}, and terabit chip-to-chip interconnects.\cite{Juan2012}

The optical spectra of WGs, however, do not consist of only these bound modes called WG modes, but also contain unbound modes which couple to the outside, commonly known as leaky modes. An elegant and intuitive way to understand and describe the properties of optical systems is to use the concept of discrete resonant states\cite{Siegert1939, Weinstein1969} (RSs) which include all types of modes in the system and present a mathematically complete set of spatial functions. RSs are defined as eigensolutions of the Maxwell equation having outgoing wave boundary conditions. Their energies are generally complex reflecting the fact that these states decay in time and leak out of the system. RSs with complex energies are therefore characterized by exponentially growing tails outside the system that require an adapted normalization.\cite{Siegert1939, Weinstein1969, Muljarov2010, Doost2012, Doost2013} WG modes, which may exist in the system, are included in the set of RSs and are required for  the completeness of the set, even though they have real energies and evanescent tails.

To calculate the RSs in optical systems in which analytic solutions are not possible, the resonant state expansion (RSE), a rigorous perturbation method in electrodynamics, has been recently developed\cite{Muljarov2010} and applied to finite 1D and 2D systems, such as planar\cite{Doost2012} and cylindrical\cite{Doost2013} resonators.
The RSE was shown to be particularly suited for the calculation of sharp resonances, such as whispering gallery modes in microcylinders\cite{Chantada2008,Dubertrand2008,Dettmann2009} and microspheres,\cite{Collot1993} where available computational techniques\cite{Taflove2000, Hagness1997,Wiersig2000, Zienkiewicz2000, Rahman1991} need prohibitively large computational resources.\cite{Jang1996,Boriskin2008}

Up to now, the RSE has been applied only to modes with zero wave vector $p$ along the translationally invariant direction of the system considered, corresponding to normal incidence of light without propagation along the waveguide. In this work, we extend the application of RSE to arbitrary wave vectors $p$, thus allowing to describe the propagation along waveguide structures. This introduces in the spectrum of RSs, which for normal incidence is dominated by lossy Fabry-Perot (FP) modes, WG and anti-waveguide (AWG) modes, as well as a continuum of modes due to a cut of the Green's function in the complex frequency plane appearing for $p \neq 0$. The modes on the cut contribute significantly to the optical spectra and are required for the completeness of the RS basis. They present a challenge in the technical implementation of the RSE which is dealing with discrete states. We have recently shown\cite{Doost2013} that one can make an effective discretization of such continua for the RSE applied to 2D systems which show a cut already for $p=0$. In the present work, we eliminate the cut in planar systems with $p\neq 0$ by going from the frequency representation of the system to the normal wave-vector representation.

We treat here planar WGs, while the application of the RSE to fiber WGs, generalizing our recent work on cylindrical resonators\cite{Doost2013} to non-normal incidence, will be the subject of a future work. We verify our theory on exactly solvable structures such as a homogeneous dielectric slab and a Bragg-mirror microcavity, using the RSs of a reference slab as a basis for the RSE. The role of the different types of RSs is studied in detail, revealing the importance of WG modes in the transmission.

The paper is organized as follows. In \Sec{Sec:Transmission} we study the transmission of a homogeneous slab in the complex frequency and normal wave vector plane, in order to analyze the contributions of different types of RSs to the optical spectra of planar WGs. In \Sec{Sec:RSE} we present a general formulation of the RSE for planar systems with non-zero in-plane momentum. In \Sec{Sec:Results} we demonstrate applications of the RSE to different systems and compare results with available exact solutions. In particular, we introduce in \Sec{Sec:Unperturbed} the basis of RSs for a homogeneous slab in inclined geometry and then use it for calculation of optical modes of a homogeneous slab with a different refractive index in \Sec{Sec:Full_width} and of a Bragg-mirror microcavity in \Sec{Sec:Microcavity}.

\section{Role of waveguide modes in transmission spectra}
\label{Sec:Transmission}

We study the role of RSs in the transmission of a dielectric slab, and in particular the influence of the WG modes on the slab transmission. The WG modes are RSs which have zero linewidth and are present in the spectrum of a planar system at non-normal incidence of the incoming light wave. We consider a dielectric slab with thickness $2a$ in vacuum, having the real dielectric constant
\begin{equation}
\varepsilon(z )=\left\{
\begin{array}{cl}
\epss & \text{for\ \  } |z| \leqslant a\,,\\
1 & \text{for\ \  } |z| > a\,,
\end{array} \right.
\end{equation}
where $\epss$ is the permittivity of the slab and $z$ is the coordinate normal to the slab. We assume a permeability of $\mu=1$  everywhere throughout this work. The electric field ${\bf E}$ satisfies Maxwell's equation,
\begin{equation}
\left[\nabla^2-\varepsilon(z)\frac{1}{c^2}\frac{\partial^2}{\partial t^2}\right]{\bf E}({\bf r},t)=0\,,
\label{ME}
\end{equation}
and Maxwell's boundary conditions on the dielectric/vacuum interfaces.
For an incoming plane monochromatic wave with the transverse-electric (TE) polarization along $\hat{y}$ ($\hat{y}$ is the unit vector along the $y$-axis) and real frequency $\omega$, the electric field in the system takes the form
\begin{equation}
{\bf E}({\bf r},t)=\hat{y} e^{-i\omega t+ipx} E(z)\,,
\label{EF}
\end{equation}
in which $p$ is the in-plane projection of the wave vector. For the component $E(z)$ of the electric field, \Eq{ME} transforms to a 1D wave equation
\begin{equation}
\left[\frac{d^2}{dz^2}-p^2+\varepsilon\left(z\right)\frac{\omega^2}{c^2}\right]E(z)=0\,.
\label{ME1D}
\end{equation}
The electric field for $z>a$ is given by the transmitted plane wave $E(z)=t(\omega) e^{ikz}E_0$ where $E_0$ is the amplitude of the incoming wave. The field transmission through the slab $t(\omega)$ has the analytic form
\begin{equation}
t(\omega)=\frac{2ikq e^{2ika}}{2ikq\cos(2qa)+(k^{2}+q^{2})\sin(2qa)}=T(k),
\label{Transm}
\end{equation}
in which
\begin{eqnarray}
k&=&\sqrt{\left(\frac{\omega}{c}\right)^2-p^2}\,,
\label{kdef}\\
q&=&\sqrt{\epsilon_s\left(\frac{\omega}{c}\right)^2-p^2}=\sqrt{\epsilon_s k^2+(\epsilon_s-1)p^2}
\label{qdef}
\end{eqnarray}
are the $z$-components of the wave vector in vacuum and dielectric, respectively.
\Eq{Transm} shows that the transmission $t(\omega)$ is a function of the real frequency $\omega$. One can also express the transmission $t(\omega)$ as a function $T(k)$ of the normal wave vector $k$, in which $k$ takes only real positive values, as dictated by the outgoing character of the transmitted wave. The wave vector $q$ inside the slab can be complex for a dielectric with dissipation and have an arbitrary sign, reflecting the fact that waves within the slab propagate in both directions. Hence the transmission is insensitive to the sign of $q$ as seen in \Eq{Transm}.

\begin{figure}[t]
\includegraphics[bb = 32 354 542 755 clip, width=\columnwidth ]{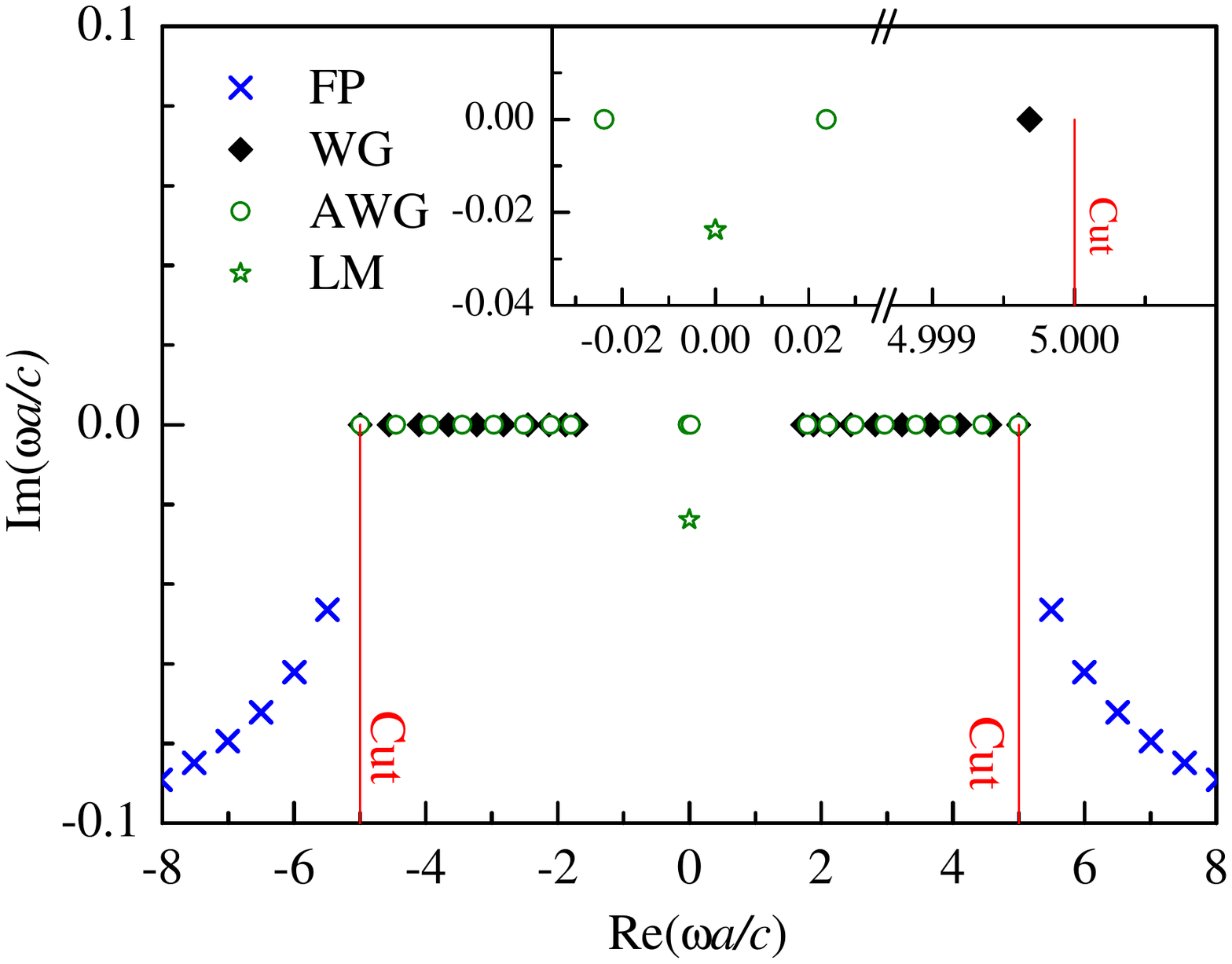}
\caption{(Color online) Poles (symbols) and cut (red lines) of the transmission $\tilde{t}(\omega)$ of a homogeneous dielectric slab with $\epss=9$ and in-plane wave vector $pa=5$. The poles are Fabry-Perot (blue crosses), waveguide (black diamonds) and anti-waveguide modes (open circles) including a leaky mode (open star). The inset shows the absence of $\omega=0$ and $k=0$ modes.} \label{fig:FIGURE1}
\end{figure}

To study the influence of different modes on the transmission, we consider analytic continuations (ACs) $\tilde{t}(\omega)$ and $\tilde{T}(k)$ of both functions in the complex $\omega$- and $k$-plane, respectively, in order to investigate their pole structure and for each of them apply the Mittag-Leffler theorem.\cite{Newton1960,More1971,M-L} The AC of the transmission has different types of poles, which are shown in \Fig{fig:FIGURE1} for $pa=5$. As in the case of normal incidence,\cite{Doost2012} there is countable infinite number of FP modes having nearly equidistant real parts and finite imaginary parts. In addition there are two types of modes on the real $\omega$-axis: WG and AWG modes, which are appearing for $p\neq0$. The WG modes have an evanescent, i.\,e. exponentially decaying electric field into the vacuum, while the AWG modes are exponentially growing into the vacuum and are known in quantum-mechanics as anti-bound states.\cite{Zavin2004}  Finally there is one leaky mode (LM) at the center of the spectrum which has zero real and negative imaginary part of $\omega$. The function $\tilde{t}(\omega)$ has two branch points at $\omega=\pm pc$ connected by a cut, due to the square root in \Eq{kdef}. We choose the cut going through $\omega=-i\infty$ and thus producing two vertical cut lines as shown in \Fig{fig:FIGURE1}. The other square root in the definition of $q(\omega)$ does not produce any cuts due to the above mentioned fact that $t(\omega)$ is an even function of $q$ and thus independent of its sign. Integrating $\tilde{t}(\omega')/(\omega-\omega')$ over a closed infinite-radius circular contour circumventing the cut, similar to that used in \Onlinecite{Doost2013}, we obtain the spectral representation in the frequency domain
\be
\tilde{t}(\omega)=\sum_n \,\frac{\res\limits_{\ \omega'=\omega_n}\left[\tilde{t}(\omega')\right]}{\omega-\omega_n}
+\frac{1}{2\pi i} \sum_{p'=\pm p}\int_{p'\!c-i\infty}^{p'\!c}\!\! \frac{\Delta t(\omega') d\omega'}{\omega-\omega'}.
\label{ML1}
\ee
Here the first term represents a sum over residues at all poles of $\tilde{t}(\omega)$. The second term is the integral of the step $\Delta t(\omega)$ in the transmission along the two parts of the cut shown in \Fig{fig:FIGURE1}. Specifically, $\Delta t(\omega)$ is defined as the difference between the values of $\tilde{t}(\omega)$ on the left and right sides of the cut for the given cut point $\omega$.

\begin{figure}[t]
\includegraphics*[bb = 45 260 455 812 clip, width=\columnwidth]{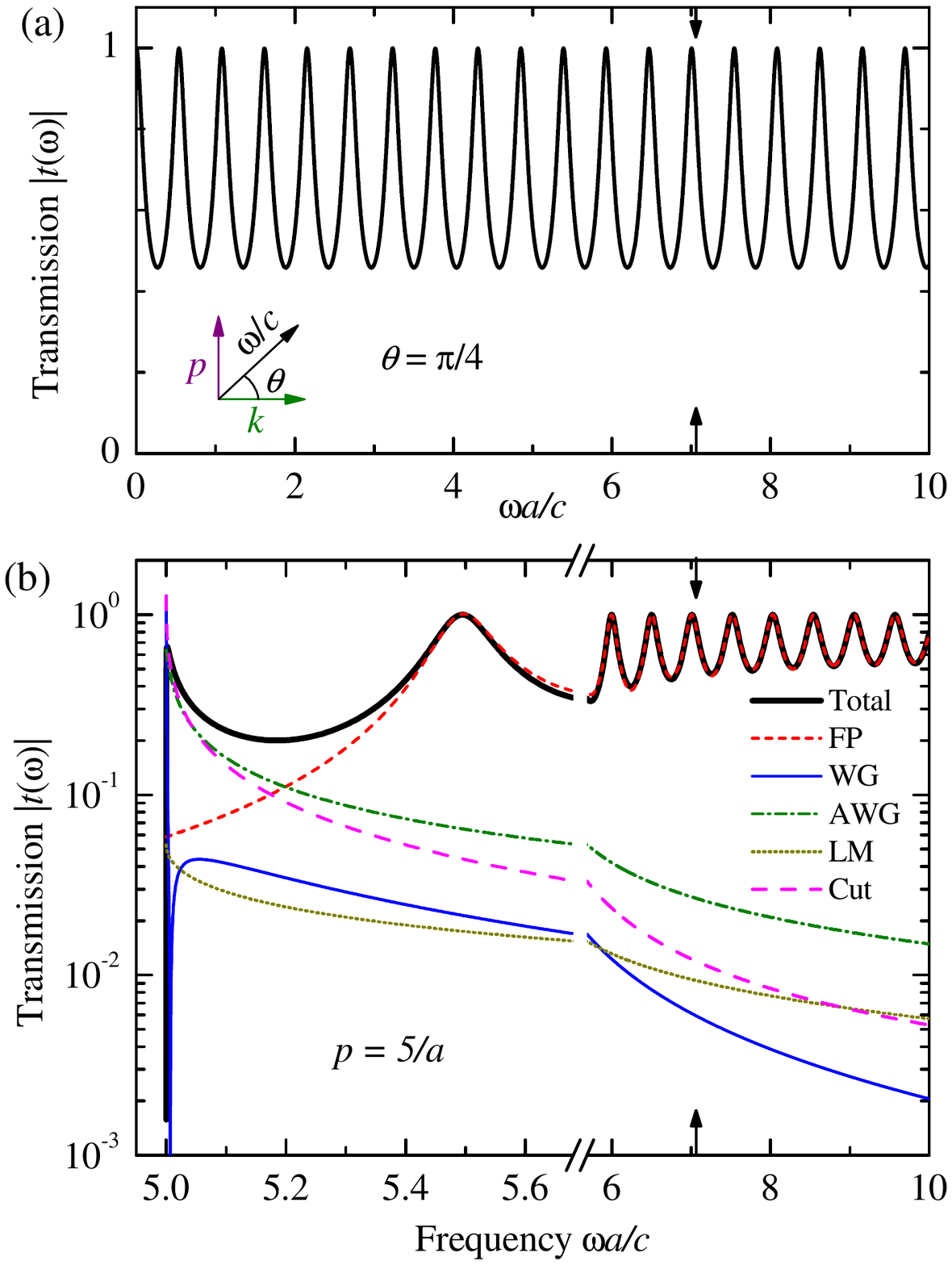}
\caption{(Color online) Transmission $|t(\omega)|$ of  a homogeneous dielectric slab with $\epsilon_s=9$ as a function of the light frequency $\omega$,  (a) for a fixed angle of incidence $\theta=\pi/4$ and (b) for a fixed in-plane wave vector  $pa=5$, along with partial contributions to the transmission of different types of modes and the cut shown in Fig.\,\ref{fig:FIGURE1}. Black vertical arrows indicate the frequency for which $pa=5$ in panel (a) and $\theta=\pi/4$ in panel  (b). The inset shows a schematic of the total wave vector $\omega/c$ along with its projections $p$ and $k$ on the $x$- and $z$-axis, respectively.
} \label{fig:FIGURE2}
\end{figure}

Using the spectral representation Eq.\,(\ref{ML1}) for real frequencies $\omega$, we analyze contributions of the poles and the cut to the transmission. The transmission is usually studied for a fixed angle of incidence $\theta$, motivated by experimental constraints. An example of the calculated transmission through a slab with $\epss=9$ is shown for $\theta=\pi/4$ in \Fig{fig:FIGURE2}\,(a).
For a fixed $\theta$, the in-plane wave vector $p$ changes with frequency, so that the contributions of the poles (which are different for different $p$) are not constant across the spectrum. We therefore analyze the spectrum for a fixed $p$, as shown in \Fig{fig:FIGURE2}\,(b), in which the contributions of different pole types and the cut are shown individually, summing up to the analytic transmission \Eq{Transm}. Note that the transmission $t(\omega)$ is defined over the angle range $0<\theta<\pi/2$, corresponding to $\omega>pc$. FP modes dominate for $\omega\gg pc$ giving rise to the oscillations in the transmission, while the contribution of all other modes and the cut are significant only close to the threshold  $\omega=pc$, corresponding to grazing incidence $\theta\sim\pi/2$.

\begin{figure}[t]
\includegraphics*[bb = 38 485 456 772 clip, width=\columnwidth]{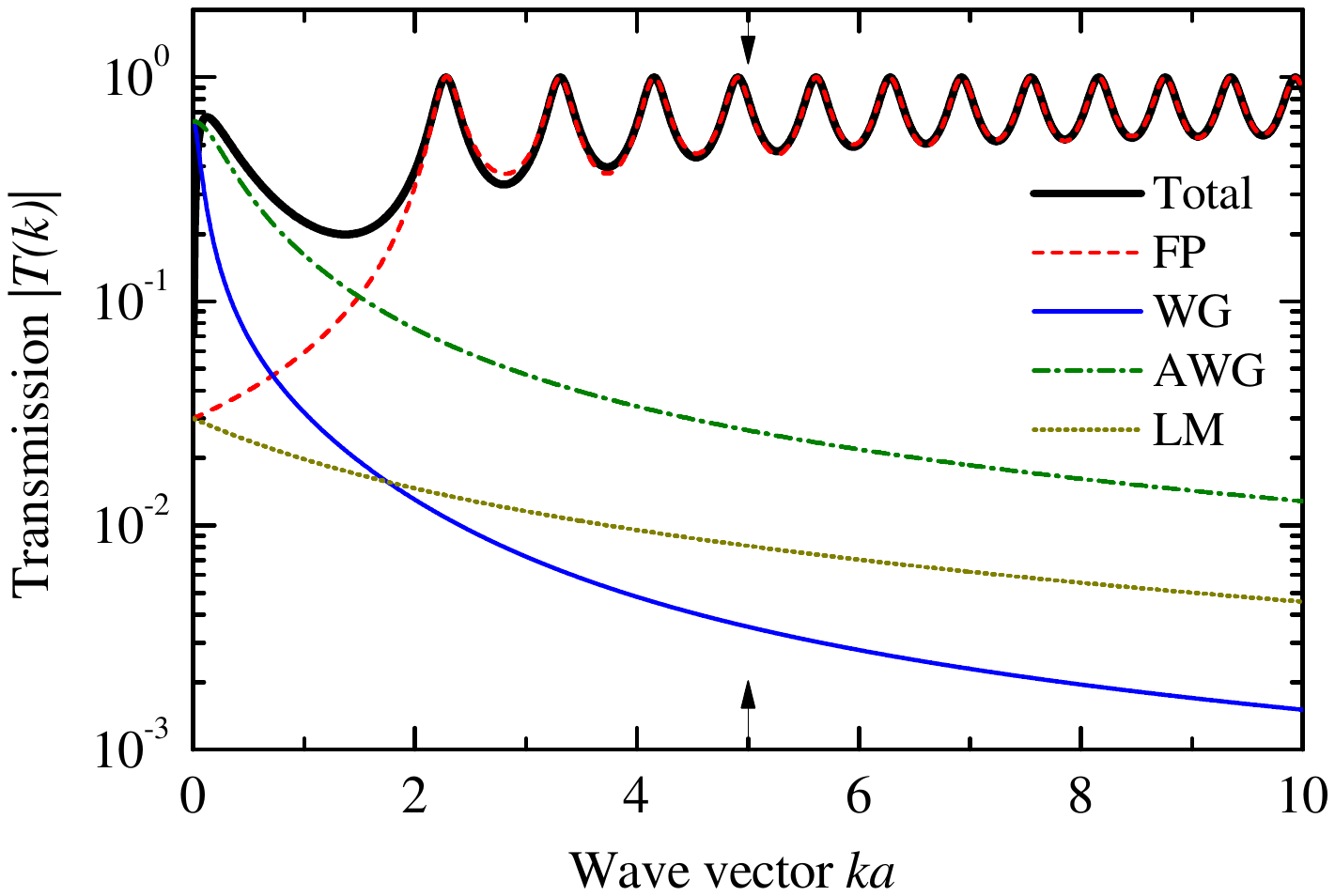}
\caption{(Color online) Transmission $|T(k)|$ of  a homogeneous dielectric slab with $\epss=9$ and $pa=5$ and partial contributions of different modes, as functions of the normal component of the wave vector in vacuum $k$. As in Fig.\,\ref{fig:FIGURE2}\,(b), vertical arrows indicate the wave vector at which  $\theta=\pi/4$.
} \label{fig:FIGURE3}
\end{figure}

The cut contribution to the spectral representation Eq.\,(\ref{ML1}) and to the transmission in Fig.\,\ref{fig:FIGURE2}\,(b) produces a continuum of resonances. Such a continuum can be approximately treated in the RSE by replacing it with a series of poles, as done in \Onlinecite{Doost2013}. In the present case however, the cut can actually be removed by going into the wave-vector domain. Indeed, being treated as a function of the normal wave vector $k$, the AC of the transmission $\tilde{T}(k)$ has no cuts in the complex $k$-plane and its spectral representation obtained by using the Mittag-Leffler theorem has the following form:
\be
\tilde{T}(k)=\sum_n \,\frac{\res\limits_{\ k'=k_n}\left[\tilde{T}(k')\right]}{k-k_n}
\,,
\label{ML2}
\ee
in which $k_n=\sqrt{\omega^2_n/c^2-p^2}$, with $n$ numbering the poles as in \Eq{ML1}. On the real $k$-axis, $\tilde{T}(k)$ coincides with the transmission ${T}(k)$ given by Eq.\,(\ref{Transm}) and is shown in Fig.\,\ref{fig:FIGURE3} along with the contributions of the different types of modes. We see in particular that the WG modes, which are not emitting into an outgoing plane wave, and thus by reciprocity are expected not to be excitable by an incoming plane wave, have a finite contribution to the transmission, which is possible only due to their off-resonant excitation.  This contribution increases with decreasing the wave vector $k$, as the frequency of the incoming wave is getting closer to the resonant frequencies of the WG modes lying beyond the vacuum light cone.

\section{Resonant state expansion for non-normal incidence}
\label{Sec:RSE}

The RSE formulated in our previous works\cite{Muljarov2010,Doost2012,Doost2013} is based on three key elements which are: (i) Dyson's equation for the Green's function (GF), (ii) spectral representation of the GF, and (iii) completeness of RSs used for expansions of perturbed states and the GF itself. We have previously applied the RSE to infinite 1D and 2D systems at normal incidence. The non-normal incidence, characterized by $p \neq 0$, is treated here. The previously used spectral representation of the GF in the frequency domain contains a cut for $p\neq 0$, which however can be removed by mapping the problem onto the complex normal wave-vector space $k$, as demonstrated in \Sec{Sec:Transmission}. We therefore reformulate the RSE in the complex $k$-plane, for which the spectral representation of the GF of an infinite planar system with an in-plane momentum $p\neq0$ for TE polarization\cite{foot2} has the form
\be
G_k(z,z')=\sum_n\frac{E_{n}(z)E_{n}(z')}{2k_{n}(k-k_{n})}\,,
\label{GF1}
\ee
where $E_n(z)$ is the electric field of a RS, defined as an eigensolution of \Eq{ME1D} with an arbitrary profile of $\varepsilon(z)$ within a finite interval $|z|<a$, satisfying the outgoing wave boundary conditions
\be
E_{n}(z) \propto e^{ik_n|z|}\ \ \ {\rm for}\ \ |z|>a
\label{BC}
\ee
and orthonormality conditions\cite{foot1}
\begin{eqnarray}
&&\int_{-a}^{a} \varepsilon(z){E}_{n} (z){E}_{m} (z)\,dz \nonumber\\
&&- \frac{{E}_{n} (-a){E}_{m} (-a) +{E}_{n} (a){E}_{m} (a)}{i({k}_{n}+{k}_{m})}=\delta_{nm}.
\label{norm}
\end{eqnarray}
The GF satisfies the equation
\begin{equation}
\left[\frac{d^2}{dz^2}-p^2+\varepsilon(z)(k^2+p^2)\right]G_k(z,z')=\delta(z-z')
\label{GFequ}
\end{equation}
and thus has the asymptotics $G_k(z,z')\propto k^{-2}$ for $k\to\infty$. Applying it to \Eq{GF1}, and using the fact that the GF has no pole at $k=0$ corresponding to $\omega=\pm pc$, as exemplified in \Fig{fig:FIGURE1}, we obtain the following sum rule:
\begin{eqnarray}
\sum_n\frac{E_{n}(z)E_{n}(z')}{k_{n}}=0\,.
\label{sum_rule}
\end{eqnarray}
For $p=0$ the right-hand side of the above sum rule is replaced by $i$, due to the $k=0$ pole of the GF.\cite{Muljarov2010}
Using \Eq{sum_rule}, one can write \Eq{GF1} as
\be
G_k(z,z')=\sum_n\frac{E_{n}(z)E_{n}(z')}{2k_{n}}\left[\frac{1}{k-k_{n}}+F(k)\right]\,,
\label{GF2}
\ee
where $F(k)$ is an arbitrary function which will be appropriately chosen later, in order to linearize a resulting matrix eigenvalue problem of the RSE.

We now consider an arbitrary perturbation $\Delta\varepsilon(z)$ of the dielectric constant inside the layer $|z|<a$.
The new, perturbed GF ${\cal G}_k(z,z')$ is related to the unperturbed one via the Dyson equation
\bea
&&{\cal G}_k(z,z')=G_k(z,z') \label{Dyson}\\
&&\ \ \ \ \ \ \ \ -(k^2+p^2)\int_{-a}^a G_k(z,z'')\Delta\varepsilon(z'') {\cal G}_k(z'',z') dz''\,. \nonumber
\eea
Substituting \Eq{GF2} into \Eq{Dyson} and using a similar spectral representation for the perturbed GF in terms of the perturbed modes ${\cal E}_\nu(z)$, we equate, following \Onlinecite{Muljarov2010}, residua at the perturbed poles $k=\varkappa_{\nu}$ in Eq.\,\eqref{Dyson}. This results in the following relationship between unperturbed and perturbed modes
\bea
{\cal E}_\nu(z)&=&-\left(\varkappa_\nu^2+p^2\right)\sum_n\frac{E_n(z)}{2k_n}\left[\frac{1}{\varkappa_\nu-k_n}+F(\varkappa_\nu)\right] \nonumber\\
&&\times \int^{a}_{-a}E_n(z')\Delta\varepsilon(z'){\cal E}_{\nu}(z')dz'\,. \label{Dyson2}
\eea
Note that the perturbed modes ${\cal E}_\nu(z)$ satisfy \Eq{ME1D} with $\varepsilon(z)$ replaced by $\varepsilon(z)+\Delta\varepsilon(z)$ and the BCs \Eq{BC} with $k_n$ replaced by $\varkappa_\nu$.
In the interior region $\left|z\right|<a$ which contains the perturbation, the perturbed RSs can be expanded into the unperturbed ones, exploiting the completeness of the latter:
\begin{equation}
\mathcal{E}_{\nu}(z)=\sum_{n}b_{n\nu}E_n(z)\,. \label{superp}
\end{equation}
Substituting this expansion into \Eq{Dyson2} and equating coefficients at the same basis functions $E_n(z)$ results in the matrix equation
\be
b_{n\nu}=-\frac{\varkappa_\nu^2+p^2}{2k_n}\left[\frac{1}{\varkappa_\nu-k_n}+F(\varkappa_{\nu})\right]\sum_m V_{nm}b_{m\nu}\,,
\label{Dyson3}
\ee
where
\begin{equation}
V_{nm}=\int_{-a}^{a}{\Delta\varepsilon(z)E_n(z)E_m(z)\,dz}
\label{Vnm}
\end{equation}
is the matrix of the perturbation in the basis of unperturbed RSs.

Equation (\ref{Dyson3})  is a matrix eigenvalue problem which can be solved numerically in order to find the wave vectors $\varkappa_\nu$ and the corresponding eigenfrequencies of the perturbed RSs, as well as their expansion coefficients $b_{n\nu}$ in terms of the unperturbed ones. However, this problem is generally nonlinear in $\varkappa_\nu$, as can be seen by choosing  $F(k)=0$. Nonlinear eigenvalue problems are known to lead to numerical instabilities and can produce spurious solutions. In order to avoid these issues, we choose
\begin{equation}
F(k)=-\frac{k}{k^2+p^2}=-\frac{kc^2}{\omega^2}\,,
\end{equation}
explicitly depending on the in-plane wave vector $p$, which linearizes the eigenvalue problem. Indeed, with the substitution $c_{n\nu}=b_{n\nu}\sqrt{k_n/\varkappa_\nu}$, the eigenvalue problem is given by
\be
\sum_m\!\!\left(\!\frac{\delta_{nm}}{k_n}\!+\!\frac{V_{nm}}{2\sqrt{k_n k_m}}\!\right)\!\!c_{m\nu}
\!=\!\frac{1}{\varkappa_{\nu}}\!\sum_{m}\!\!\left(\!\delta_{nm}\!-\!\frac{p^2 V_{nm}}{2k_n\sqrt{k_n k_m}}\!\right)\!\!c_{m\nu}
\label{Dyson4}
\ee
which is {\it linear} and can be solved by inverting the matrix on the right-hand side of \Eq{Dyson4} and diagonalizing the resulting non-symmetric matrix on the left-hand side, in order to obtain its eigenvalues $1/\varkappa_\nu$. Alternatively, one can solve \Eq{Dyson4} by employing a variety of software libraries available for generalized linear matrix eigenvalue problems. Note that the matrix equation of the RSE for normal incidence previously derived in \Onlinecite{Muljarov2010} is restored by choosing $p=0$ in \Eq{Dyson4}.

\section{Results}
\label{Sec:Results}

To apply the method developed in \Sec{Sec:RSE} we first construct a convenient basis of unperturbed states. We use the RSs of a homogeneous dielectric slab discussed in \Sec{Sec:Transmission}. We calculate the wave functions $E_n(z)$ of the RSs and investigate the dependence of their eigenvalues $k_n$ on the in-plane wave vector $p$. Then, using the RSE, in particular Eqs.\,(\ref{Vnm}) and (\ref{Dyson4}), we calculate the perturbed eigenvalues $\varkappa_\nu$ for the simplest perturbation, which is constant across the slab, and compare the RSE results with the available exact solution. Finally, we use the RSE to treat a structured perturbation simulating a Bragg-mirror microcavity (MC). We specifically discuss the lowest-energy cavity mode (CM) and compare results with the transfer matrix calculation of the MC transmission and with an available analytic approximation for the CM linewidth.

\subsection{Unperturbed resonant states}
\label{Sec:Unperturbed}

The solutions of \Eq{ME1D} which satisfy the outgoing-wave boundary conditions \Eq{BC} in TE polarization take the form
\begin{equation}
E_n(z)=\left\{
\begin{array}{lll}
(-1)^nA_ne^{-ik_nz}\,, & & z\,\leqslant-a\,,\\
B_n[e^{iq_nz}+(-1)^ne^{-iq_nz}]\,, &  &\!\!|z|\leqslant a\,,\\
A_ne^{ik_nz}\,, && z\,\geqslant a\,,
\end{array} \right.
\label{En}
\end{equation}
where the eigenvalues $k_n$ satisfy the secular equation
\begin{equation}
\left(k_n-q_n\right)e^{iq_na}+\left(-1\right)^n\left(k_n+q_n\right)e^{-iq_na}=0\,,
\label{secular}
\end{equation}
with $q_n=\sqrt{\epsilon_s k_n^2+(\epsilon_s-1)p^2}$.
We use here an integer index $n$ which takes even (odd) values for symmetric (anti-symmetric) RSs, respectively.
The normalization constants $A_n$ and $B_n$ are found from the continuity of $E_n$ across the boundaries and the normalization condition \Eq{norm}. They take the form
\bea
A_n&=&\frac{e^{-ik_na}}{\sqrt{a(\epss-1)}}\sqrt{\frac{\epss \omega_n^2/c^2-p^2}{\epsilon_s \omega_n^2/c^2+ip^2/(k_na)}}\,,
\label{An}
\\
B_n&=&\frac{(-i)^n}{2\sqrt{a\epss+ip^2/(k_n \omega_n^2/c^2)}}\,,
\label{Bn}
\eea
where $\omega_n^2/c^2=k_n^2+p^2$.

\begin{figure}[t!]
\includegraphics*[bb = 30 13 410 818 clip, width=0.9\columnwidth]{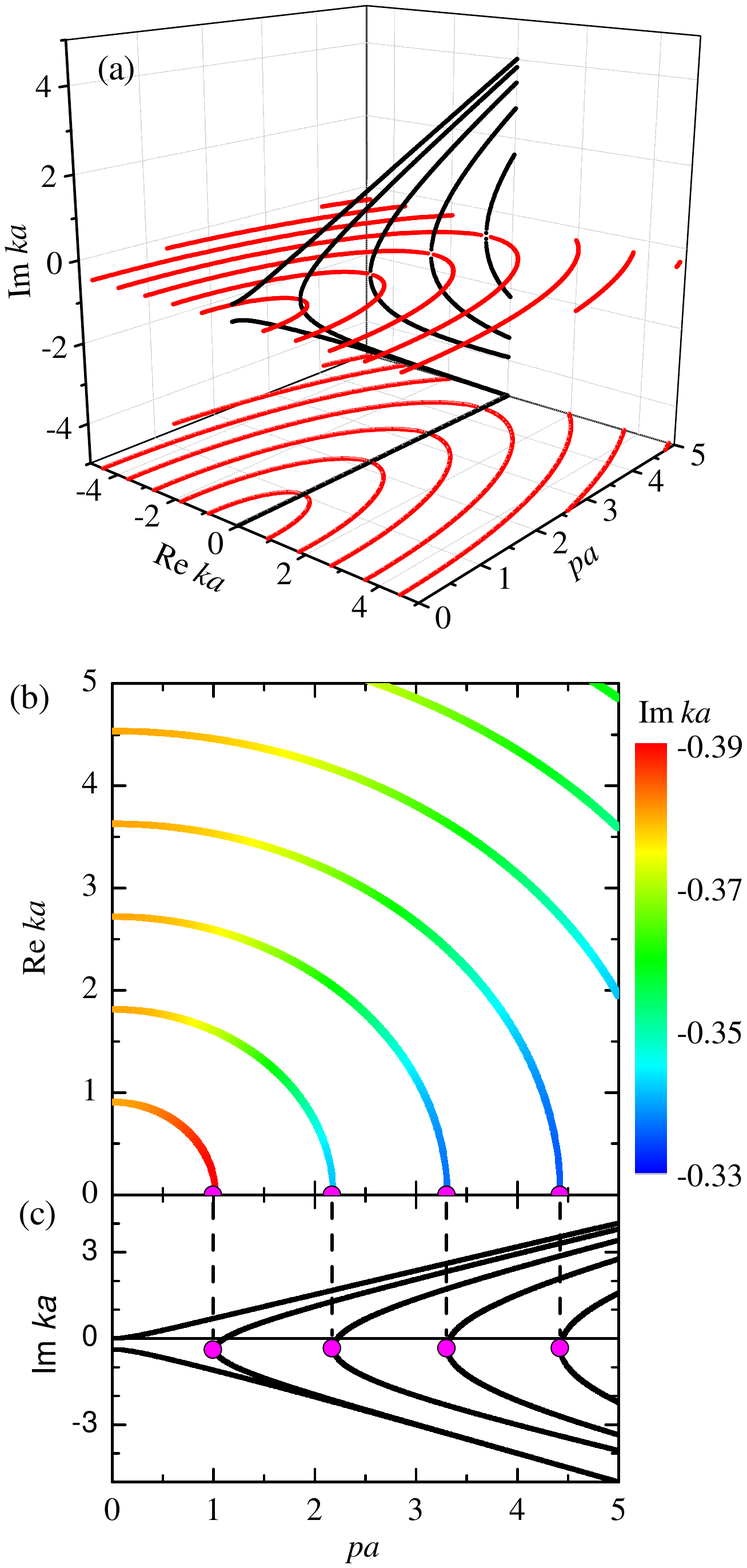}
\caption{(Color online) Resonant state wave numbers of a homogenous dielectric slab with $\epss=3$ as function of the in-plane wave vector $p$: (a) The complex wave vectors $k_n$ of Fabry-Perot (red lines) and WG and AWG modes (black lines), with a projection on the lower plane; (b) ${\rm Re}\, k_n$ of Fabry-Perot modes with the color giving the value of ${\rm Im}\, k_n$; (c) ${\rm Im}\, k_n$ of the WG and AWG modes. The points where the modes in panels (b) and (c) connect are given by magenta dots joined by dashed lines. 
}
\label{fig:FIGURE4}
\end{figure}

The frequencies $\omega_n$ of the RSs of a dielectric slab for $pa=5$ and $\epss=9$ were shown in \Fig{fig:FIGURE1}. The normal wave vectors $k_n$ of the RSs for a slab with $\epss=3$ versus $p$ are given in \Fig{fig:FIGURE4}. All states in the range
$|{\rm Re}\, k_n a|<5$ and $|{\rm Im}\, k_n a|<5$ for $|pa|<5$ are shown in \Fig{fig:FIGURE4}\,(a) and separated into mode types in \Fig{fig:FIGURE4}\,(b) and (c). For WG and AWG modes ${\rm Re}\,k_n=0$,
therefore \Fig{fig:FIGURE4}\,(c) shows only their imaginary part, which is positive for WG modes, corresponding to evanescent waves, and negative for AWG modes, corresponding to exponentially growing waves outside the slab. The WG and AWG modes continuously transform into each other and produce branches similar to those seen also for FP modes. These branches cross each other at certain points [shown in Figs.\,\ref{fig:FIGURE4}\,(b) and (c) by magenta dots] where two FP modes are transformed into two AWG modes. The AWG mode branch which starts at $p=0$ has no connection to any WG or FP branches; a mode on this branch was identified in Fig.\,\ref{fig:FIGURE1} as the leaky mode.

The RSs of the homogeneous slab shown, similar to those shown in Fig.\,\ref{fig:FIGURE4}, are used as a basis for the RSE in the two examples given below. In general, for any local perturbation $\Delta\varepsilon(z)$ which does not change the translational symmetry of the slab, i.\,e. does not depend on $x$ of $y$, the in-plane momentum $p$ remains a good quantum number. In other words, $\Delta\varepsilon(z)$ does not mix states with different $p$, so that in any such problem solved by the RSE, one can use the basis of RSs with a given fixed value of $p$.

\subsection{Full-width perturbation}
\label{Sec:Full_width}

To illustrate the accuracy and convergence of the RSE, we consider a homogeneous full-width perturbation of the slab, which is given by
\begin{equation}
\Delta\varepsilon(z)=\left\{
\begin{array}{cl}
\Delta\epsilon & \text{for\ \  } \left|z\right|\leqslant a\,,\\
0 &\text{otherwise}\,,
\end{array} \right.
\label{Pert}
\end{equation}
and for which the exact solution can be obtained by solving the  transcendental \Eq{secular} with
$\epss$ replaced by $\epss+\Delta\epsilon$. We denote these exact perturbed wave numbers as $\varkappa^{\rm(exact)}_\nu$ and compare them with the perturbed values $\varkappa_\nu$ obtained by using the RSE for different basis sizes $N$.
We choose as basis of given size all poles with $\left|k_n\right|<k_{\rm max}(N)$, using a suitably chosen  wave-number cutoff $k_{\rm max}(N)$.

\begin{figure}[t]
\includegraphics*[bb = 38 123 518 773 clip, width=\columnwidth]{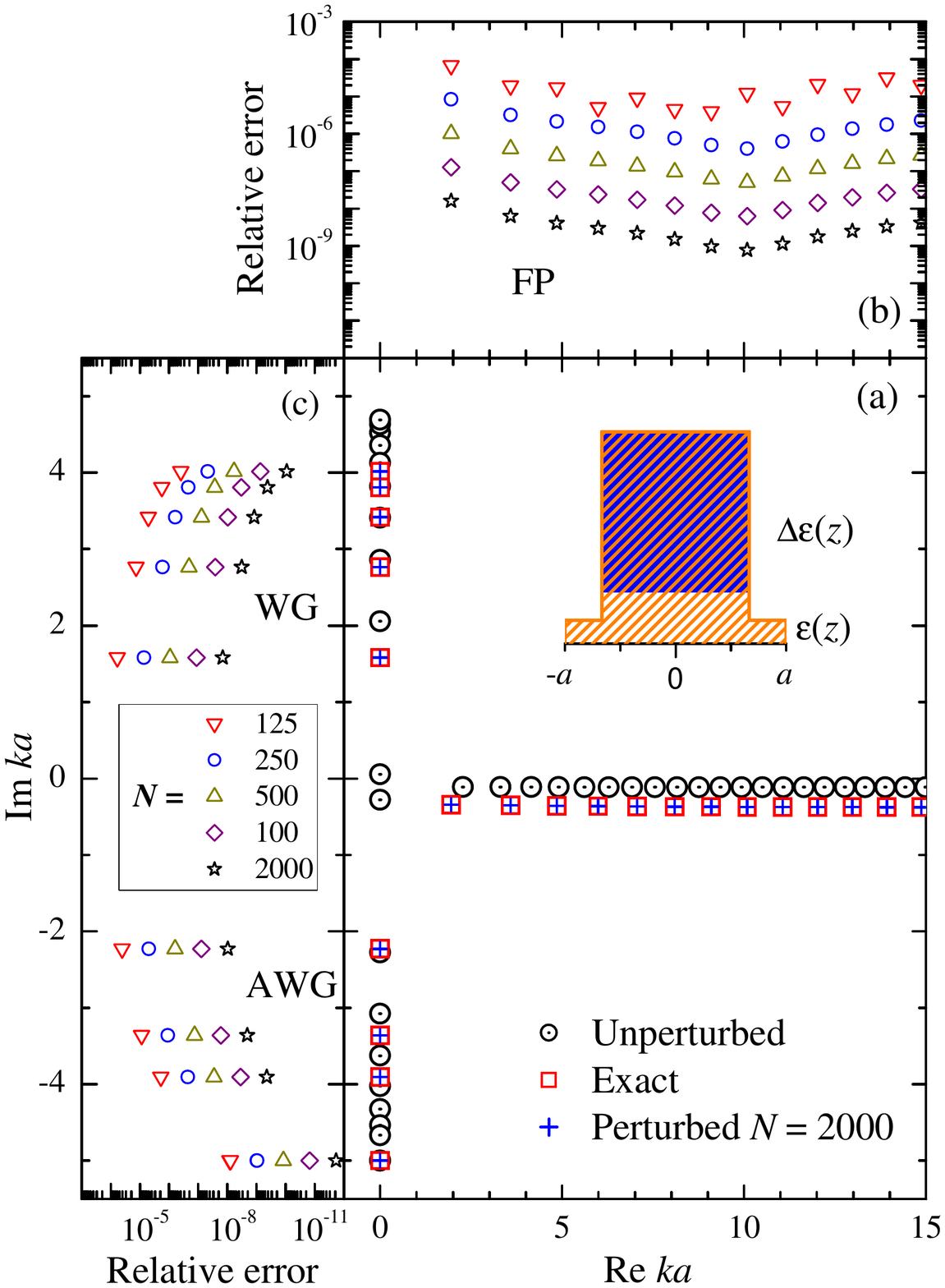}
\caption{(Color online) (a) Exact (squares) and calculated by the RSE with $N=2000$ (crosses) resonant state wave numbers of a homogeneous dielectric slab with $\epss=3$ along with those of the unperturbed slab with $\epss=9$ (circles with a dot). Relative errors in calculation of Fabry-Perot modes (b) and waveguide and anti-waveguide modes (c) for different total number of basis states $N$ used in the RSE as labeled. Inset: the dielectric constant profile of the unperturbed and perturbed systems, with the full-width homogeneous perturbation of the slab $\Delta\epsilon=-6$.} \label{fig:FIGURE5}
\end{figure}

In \Fig{fig:FIGURE5} we compare the RSE wave numbers with the exact wave numbers for our system in the case of $pa=5$. We can see in \Fig{fig:FIGURE5}(a) that the RSE is reproducing the exact solution to a high accuracy, which is quantified by the relative error $|\varkappa_\nu/\varkappa^{\rm (exact)}_\nu-1|$ shown in \Fig{fig:FIGURE5}(b) for the FP modes with ${\rm Re}\, \varkappa_\nu>0$ and in \Fig{fig:FIGURE5}\,(c) for the WG and AWG modes. We see that the relative error scales as $N^{-3}$, which was also observed in planar systems at normal incidence\cite{Muljarov2010, Doost2012} and in homogeneous micro-cylinders\cite{Doost2013} and microspheres.\cite{Muljarov2010} We find in the simulation used to generate \Fig{fig:FIGURE5} for a basis of $N=2000$ that the RSE reproduces about 300 modes with a relative error below $10^{-8}$. This error can be further improved by 1-2 orders of magnitude using the extrapolation method described in \Onlinecite{Doost2012}.

\subsection{Microcavity perturbation}
\label{Sec:Microcavity}

\begin{figure}[t]
\includegraphics*[bb = 26 201 467 774 clip,width=\columnwidth]{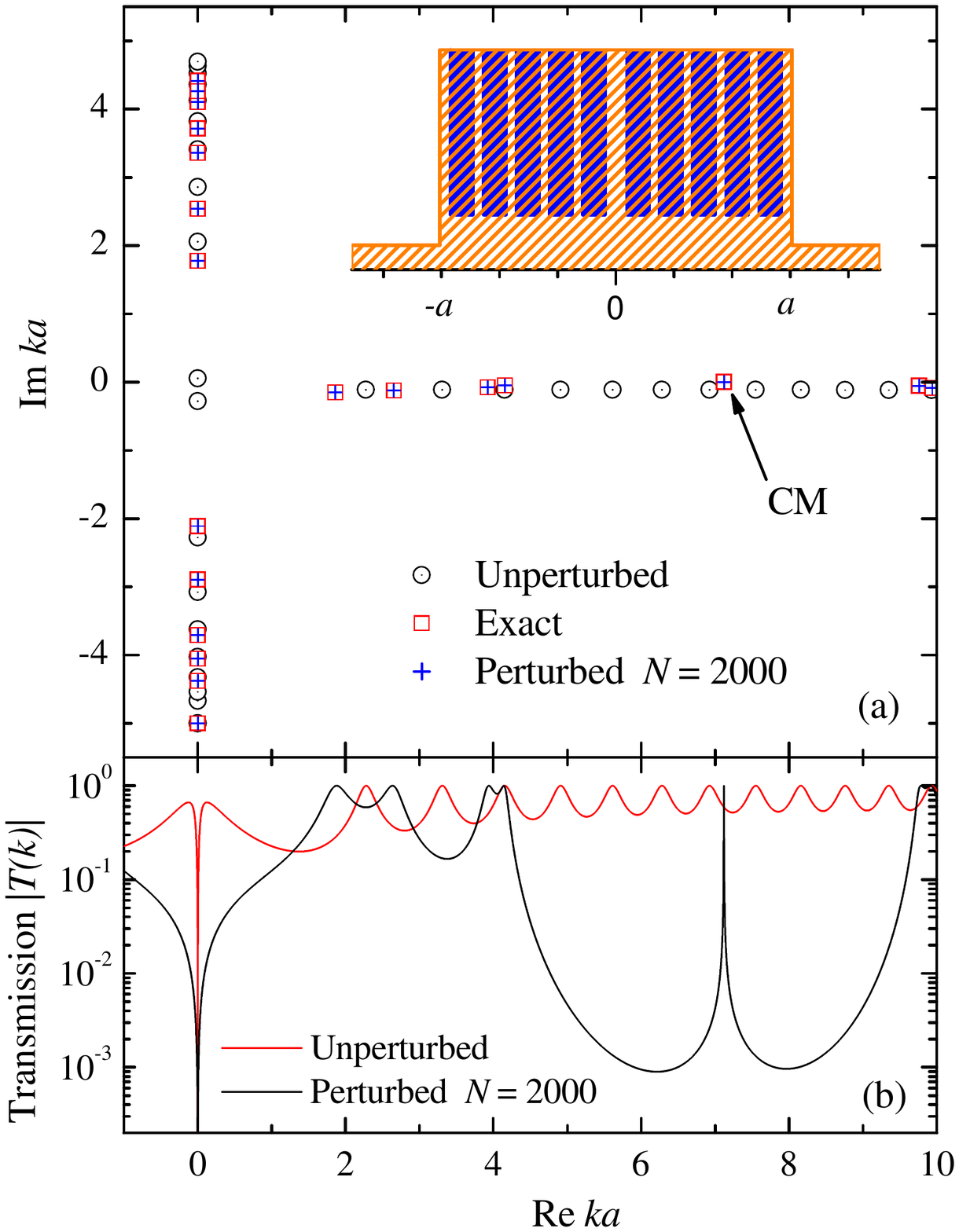}
\caption{(Color online) (a) The same as in Fig.\,\ref{fig:FIGURE5}\,(a) but with the perturbed system being the Bragg-mirror microcavity with the dielectric profile shown in the inset. The lowest-energy cavity mode is shown by an arrow. (b) Transmission as a function of the normal component of the wave vector $k$, for the perturbed (thick black curve) and unperturbed system (thin red curve) demonstrating the correspondence between the RS wave numbers in panel (a) and peaks in the transmission. }
\label{fig:FIGURE6}
\end{figure}

To evaluate the RSE for inclined geometry in the presence of sharp resonances in the optical spectrum, we use a Bragg-mirror MC, which consists of a FP cavity of thickness $L_{\rm C}$ and dielectric constant  $\epsC=9$ surrounded by distributed Bragg reflectors (DBRs). The DBRs consist of $P=5$ pairs of dielectric layers with alternating high $\epsH=9$ and low $\epsL=2.25$ susceptibility, as illustrated by the inset in Fig.\,\ref{fig:FIGURE6}. The alternating layers have a quarter-wavelength optical thickness and the cavity has a half-wavelength optical thickness. The nominal wavelength which determines the layer thickness is  that of the lowest-frequency CM at normal incidence.  As unperturbed system we used a dielectric slab with $\epss=9$ as in \Sec{Sec:Full_width}.

The unperturbed modes of the slab and the perturbed modes of the MC are shown in \Fig{fig:FIGURE6}\,(a) for $pa=5$. One can see how the nearly equidistant FP modes of the unperturbed system are redistributed in the MC, transforming into a sharp CM in the middle of a wide stop-band and modes outside of the stop-band.
The link between the peaks in the transmission in  \Fig{fig:FIGURE6}\,(b) and the poles in \Fig{fig:FIGURE6}\,(a) is also exemplified by the real part of the poles giving the position of the peaks in transmission and the imaginary part giving their linewidth. This is discussed in \Onlinecite{Doost2012} in terms of the GF which is related to transmission via $T(k)=2ik G_k(a,-a)$.

\begin{figure}[t]
\includegraphics*[bb = 23 370 482 774 clip,width=\columnwidth]{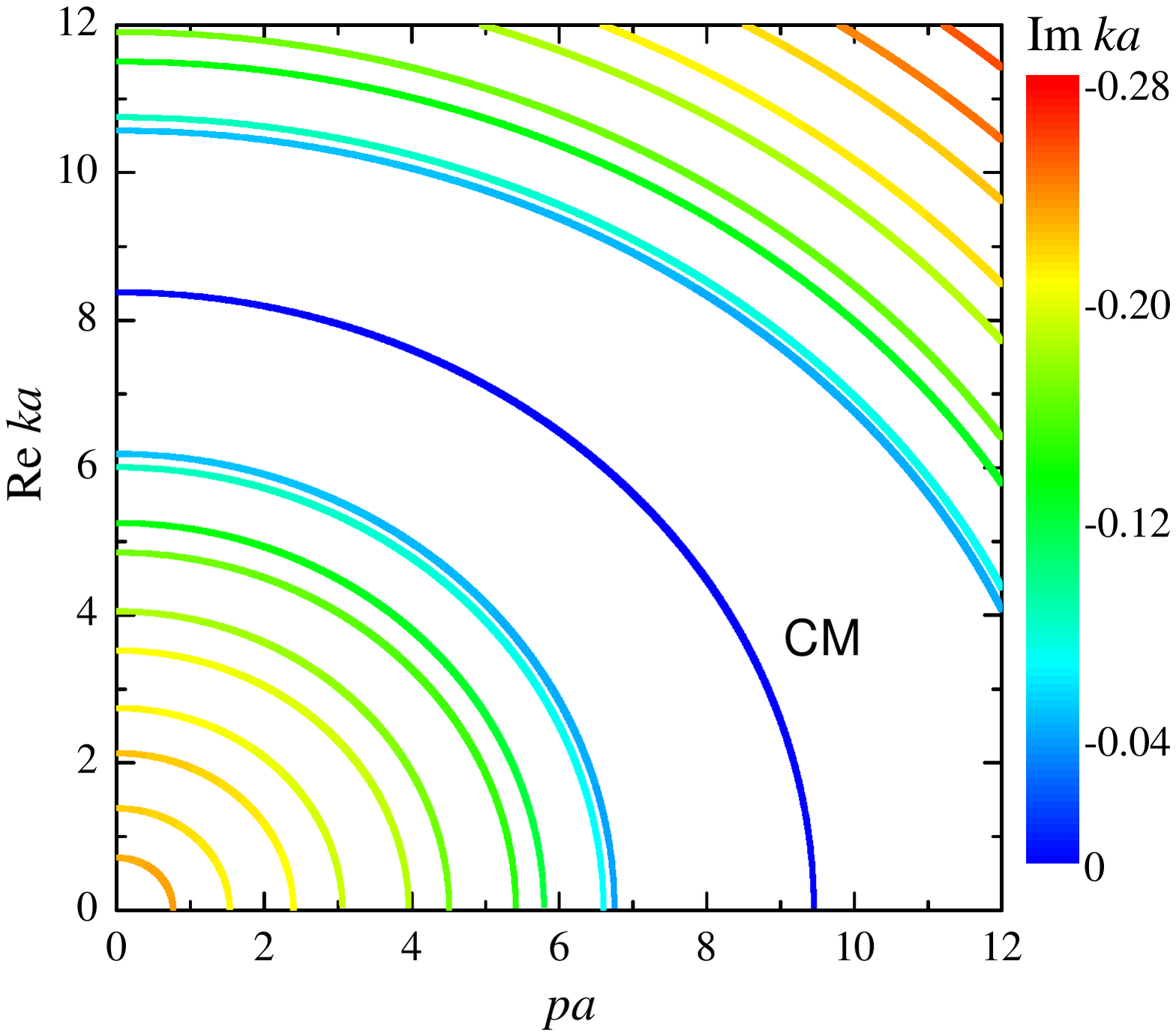}
\caption{(Color online)  The same as in \Fig{fig:FIGURE4}\,(b) but for the Bragg-mirror microcavity with the dielectric profile given by the inset in \Fig{fig:FIGURE6}\,(a). }
\label{fig:FIGURE7}
\end{figure}

The transmission $T(k)$ for a layered planar structure can be calculated using the transfer matrix method leading to the following explicit result:
\begin{equation}
T(k)=\frac{e^{i(q_0+q_M)a}}{\xi^+_M}\,,
\label{transmission}
\end{equation}
in which $\xi^+_M$ is found from the recursive formula
\be
2\xi^\pm_{j+1}=\left(1\pm\frac{q_{j+1}}{q_j}\right)e^{-iq_ja_j}\xi^+_j
+\left(1\mp\frac{q_{j+1}}{q_j}\right)e^{iq_ja_j}\xi^-_j
\label{xij}
\ee
with the starting value
\be
2\xi^\pm_1=\left(1\pm\frac{q_1}{q_0}\right)
\label{xi1}
\ee
and the normal component of the wave vector in the $j$-th layer
\be
q_j=\sqrt{\epsilon_j k^2+(\epsilon_j-1)p^2}.
\label{qj}
\ee
Here $\epsilon_j$ and $a_j$ are, respectively, the dielectric constant of the $j$-th layer and its width, so that $\sum_{j=1}^{M-1} a_j=2a$. The layers $j=0$ and $j=M$ correspond to the vacuum before and after the MC, respectively, so that $q_0=q_M=k$, and $q_j\geqslant 0$ for real $\epsilon_j$. $M$ gives the total number of interfaces in the structure, in the present case $M=2(2P+1)$.

In \Fig{fig:FIGURE7} we show the evolution of the perturbed poles with $p$. We see that one of the modes is separated in the middle of a gap and has an imaginary part well below the others. This mode is know as the CM. The perturbed Green's function ${\cal G}_k(z,z')$ which has a spectral representation equivalent to \Eq{GF1} and the corresponding transmission $T(k)$ are dominated by the single term from the CM in this frequency region, therefore a sharp isolated peak is seen in the center of the stop-band in \Fig{fig:FIGURE6}\,(b). Interestingly, the modes in \Fig{fig:FIGURE7} show an almost circular behavior, indicating that the frequency of each mode $\omega_\nu=c\sqrt{\varkappa_\nu^2+p^2}$ is approximately constant versus angle $\theta$.

\begin{figure}[t]
\includegraphics*[bb = 12 114 476 782 clip,width=\columnwidth]{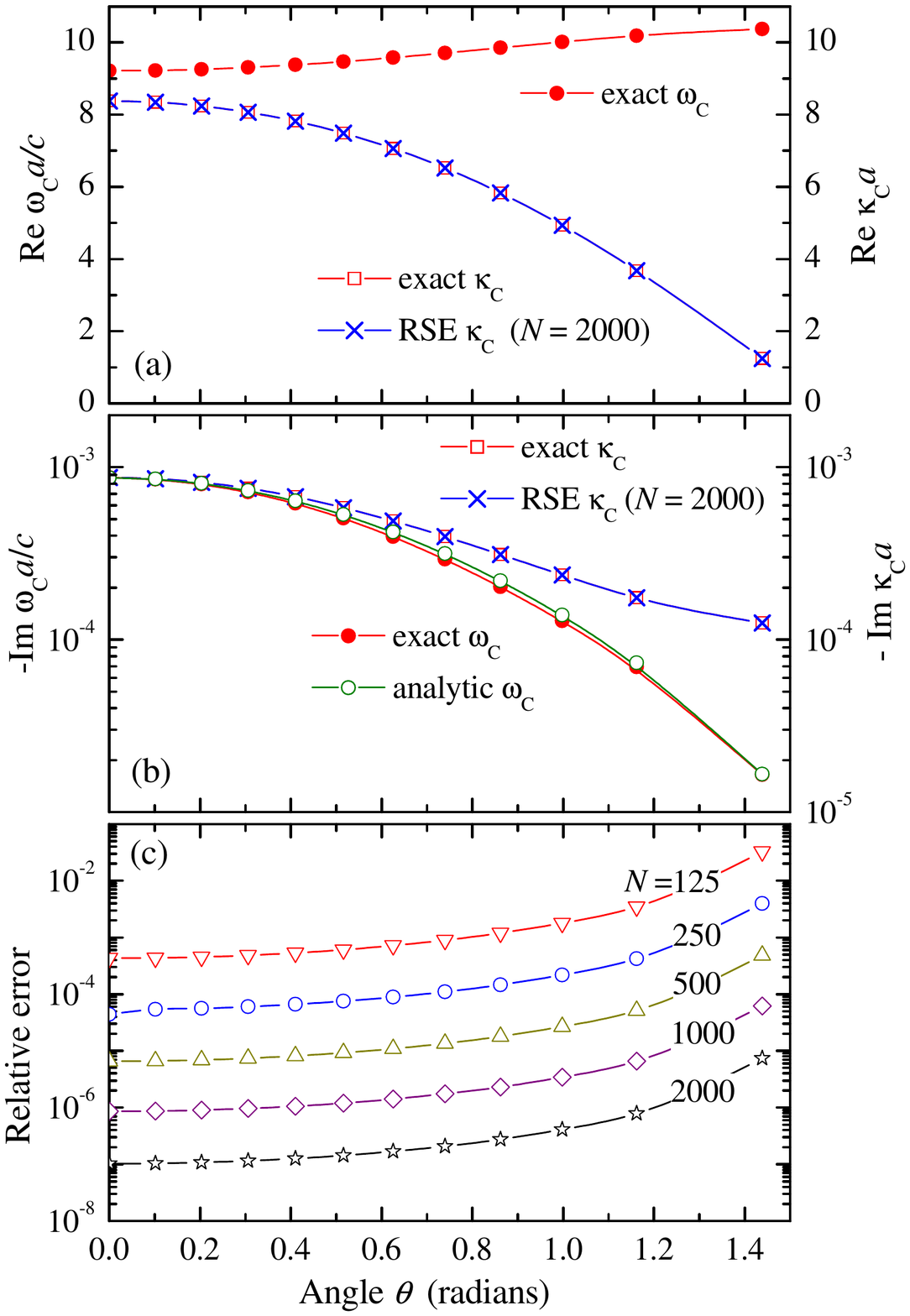}
\caption{(Color online) Real (a) and imaginary part (b) of the cavity mode frequency $\omC$ (left axes) and normal component of the wave vector $\kappaC$ (right axes) calculated using the RSE (blue crosses) for $N=2000$, the transfer matrix method (red circles and open squares) and the analytic approximation \Eq{Approx} for the linewidth $\Gamma=-{\rm Im}\,\omC$ (green open circles). (c) Relative error of $\kappaC$ determined by RSE  for different basis sizes $N$ as given. All data are shown as a function of the angle of incidence $\theta$, and all symbols are connected by lines as a guide to the eye.
 }
\label{fig:FIGURE8}
\end{figure}

Indeed, we can see in Fig.\,\ref{fig:FIGURE8}\,(a) that the CM frequency $\omC$ has a  weak dependence on $\theta$, while the corresponding wave vector $\kappaC$ changes more strongly. In parallel, the linewidth given in  \Fig{fig:FIGURE8}\,(b) shows a similar behavior both in the $\omega$- and $k$-representations, though at $\theta\to\pi/2$ the imaginary part of $\omC$ is one order of magnitude smaller than that of $\kappaC$.

Figures~\ref{fig:FIGURE8}(a) and (b) demonstrate a good agreement between $\kappaC$ obtained using the RSE and $\varkappa^{(\rm exact)}_{\rm C}$ extracted from the linewidth in the transmission calculated via Eqs.\,(\ref{transmission})--(\ref{qj}). \Fig{fig:FIGURE8}(c) shows the relative error $|\kappaC/\varkappa^{\rm(exact)}_{\rm C}-1|$ for different values of $N$ demonstrating convergence of the RSE for the cavity mode with $N^{-3}$, the same as for the homogeneous perturbation of the slab. The convergence behavior depends on the distribution of the perturbation in the wave-vector space as discussed in Refs.~\onlinecite{Doost2012} and \onlinecite{Doost2013}. Interestingly, the RSE can reproduce sharp resonances in the transmission profile, in spite of the absence of sharp resonances in the basis.

We can also compare the results in \Fig{fig:FIGURE8}(b) with an analytic approximation for the CM linewidth
\begin{equation}
{\rm Im}\, \omC=- \frac{2c\eta_{\rm ext}}{\nC\etaC}\frac{\left(\etaL/\etaH\right)^{2P}}{\LC\cos(\thetaC) + \frac{\lambdaC}{2}\frac{\etaL\etaH}{\etaH-\etaL}\frac{1}{\etaC}}\,,
\label{Approx}
\end{equation}
which we have derived by generalizing the approximation for normal incidence of light available in the literature.\cite{Andreani1994,Savona1995,Doost2012} Here $n_j$ is the refractive index of layer $j$, $\eta_{j}=n_j\cos(\theta_j)$, and $\theta_j$ is the angle to the normal in layer $j$, given by $\sin(\theta_{j})n_j=\sin(\theta)$. The layers $j$ used are: the external region (ext) which is vacuum in our case, the high-index (H) layer, the low-index (L) layer of the Bragg mirror, and the cavity layer (C). The cavity wavelength is given by $\lambdaC=2\LC\cos(\thetaC)$. Equation~(\ref{Approx}) is exact in the limit $P\to\infty$, for a structure with Bragg-mirror layer widths strictly equal to a quarter-wavelength and the cavity layer width to a half-wavelength optical thickness. This condition depends on the incident angle, and in our fixed structure is fulfilled for normal incidence only. Nevertheless, \Eq{Approx} reproduces the exact result reasonably well over the whole angle range, as shown in \Fig{fig:FIGURE8}(b).

\section{Conclusion}
\label{Sec:Conclusion}

We have generalized the resonant state expansion to planar optical systems with inclined geometry. The method is based on the spectral representation of the Green's function of Maxwell's equation and expansion of the optical modes of a perturbed system into a complete set of resonant states of a simple dielectric slab. In inclined geometry, the spectrum of a planar system contains a continuum of resonances originating from a cut of the Green's function, which we have eliminated by mapping the frequency into the normal wave vector. The optical modes and spectra of a perturbed planar system are then calculated by solving a linear matrix eigenvalue problem containing matrix elements of the perturbation in the basis of discrete resonant states only. We have verified the method on full-width homogeneous and Bragg-mirror microcavity perturbations and compared results with obtained analytic solutions, demonstrating fast convergence of the method towards the exact result. We have recently demonstrated the application of RSE to two-dimensional open optical systems at normal incidence. We expect that we can extend this treatment to inclined geometry using a similar approach, which would provide an efficient algorithm to calculate the optical modes in fibers and waveguides, including photonic crystal fibers having a complex structure.

\acknowledgments M.D. acknowledges support by the EPSRC under the DTA scheme.

\end{document}